\documentclass[doublecol]{epl2}
\usepackage{amsmath}
\usepackage{amsfonts}
\usepackage{graphicx}
\usepackage{amssymb}
\usepackage{float}

\title{Fano resonances induced by the topological magnetoelectric\,effect}

\author{L. A. Castro-Enriquez \inst{1} \and A. Mart\'{i}n-Ruiz \inst{2}} \institute{ \inst{1} Programa de F\'{i}sica, Universidad del Quind\'{i}o, 630001 Armenia,  Colombia. \\ \inst{2} Instituto de Ciencias Nucleares, Universidad Nacional Aut\'{o}noma de M\'{e}xico, 04510 Ciudad de M\'{e}xico, M\'{e}xico}


\abstract{We investigate the interaction between a topological insulator nanoparticle and a quantum dot in an impulse magnetic field. Since topological insulators are nonmagnetic, after the impulse has ended only the localised topological surface modes, which are quantised in terms of dipolar bosonic modes, thus coupling dipolarly to the quantum dot. Hence, the hybrid system can be treated as a single bosonic mode interacting with a two-level system, where the coupling strength is quantised in terms of the magnetoelectric polarizability. We implement the interaction of the hybrid with the environment through the coupling with a continuum reservoir of radiative output modes and a reservoir of phonon modes. Using the method of Zubarev’s Green functions, we derive an expression for the optical absorption spectrum of the system. We find the emergence of Fano resonances which are direct manifestations of the $\mathbb{Z}_{2}$ invariant of topological insulators. We present numerical results for a topological insulator nanosphere made of TlBiSe$_2$ interacting with a CdSe quantum dot.}

\begin{document}

\maketitle

\section{I. Introduction} 
Topological insulators (TIs) are an interesting class of materials that behave as insulators in the bulk while exhibiting conducting helical surface or edge states that are protected by time-reversal (TR) symmetry \cite{TI-Qi, TI-Hasan}. Aside from their microscopic distinguishing features, TIs also exhibit bulk magnetoelectricity when TR symmetry is broken on the surface \cite{Qi-TFT}. In TIs, the application of a magnetic field induces an electric polarisation, whereas an electric field induces a magnetisation. This is the topological magnetoelectric (TME) effect, which is characterised by a coefficient quantised to the fine structure constant. The experimental observation of the quantised magnetoelectric response is specially important because it would corresponds to a direct measurement of the $\mathbb{Z}_{2}$ invariant through a response function.

The optical properties of hybrid plexcitons (i.e. systems formed by semiconductor quantum dots and plasmonic nanostructures) have attracted great attention due to its potential applications in photonics and optoelectronics. One of the distinguishing features of hybrid plexcitons is the emergence of Fano resonances in the optical absorption spectra \cite{Miroshnichenko}. While originally developed to explain the inelastic scattering of electrons in helium \cite{Madden}, the ultrasharp spectrum of Fano resonances has become itself in a centerpiece in the realization of many modern optical devices \cite{Zhao, Cao, Heeg}. Also, its peculiar asymmetric line shape is found to be extremely sensitive to environmental changes, a characteristic which has enabled the realization of highly sensitive and accurate sensors \cite{Wu, Li}. In this Letter we suggest that such sensitiveness, together with the recent advances in the fabrication of nanostructured devices made from TI materials (i.e. TI nanoparticles \cite{Jia, Claro} and nanowires \cite{Jauregui, Siroki}), can be used to form a {\it topological hybrid plexciton} from which the $\mathbb{Z}_{2}$ invariant of TIs can be measured.

The system we shall consider is composed by a quantum dot interacting with a TI nanoparticle subject to an applied magnetic field. As we discuss now, the appearance of Fano resonances in this hybrid will be a direct signature of the $\mathbb{Z}_{2}$ invariant of TIs. Consider a magnetically permeable sphere subject to a uniform magnetic field. The magnetic field will magnetise the material, which can be described by the appearance of an image magnetic dipole at the center of the sphere. If the same thing is done with a topological insulator sphere, of which the surface states have been gapped by TR symmetry breaking, in addition to the image magnetic dipole an image electric dipole will also appear at the same place. Since TIs are mostly nonmagnetic, the induced dipoles will be topological in nature, i.e. they are direct manifestations of the TME effect. Therefore, this configuration is appropriate to measure the $\mathbb{Z}_{2}$ invariant of TIs. To this end, in this Letter we investigate the effect of an impulse magnetic field upon the topological insulator-quantum dot hybrid. Once the impulse has ended, only the localised topological modes will remain at the TI surface, which can be described by well-defined dipolar bosonic modes, and which in turn couple dipolarly to the intrinsic electric dipole moment of the quantum dot. Taking into account the coupling with the environment (through the coupling with a continuum reservoir of radiative output modes and a reservoir of phonon modes), we establish the Hamiltonian describing the full open quantum system and show, by using the Zubarev's Green function method, that Fano resonances appear in the optical absorption spectrum. These {\it topological Fano resonances} are direct manifestation of the $\mathbb{Z}_{2}$ invariant of TIs. For numerical calculations we consider the specific case of a Cadmium Selenide (CdSe) QD in proximity to a topological insulator nanoparticle made of TlBiSe$_{2}$, both immersed in a polymer layer such as poly(methyl methacrylate).

\section{II. Classical electrodynamics considerations} \

{\it Topological magnetoelectric effect.} The electromagnetic response of a conventional phase of matter (namely, insulators and metals) is governed by Maxwell equations derived from the ordinary electromagnetic Lagrangian $\mathcal{L} _{0} = (1/8 \pi) [ \epsilon \epsilon _{0} \boldsymbol{E} ^{2} - (1/\mu \mu _{0}) \boldsymbol{B} ^{2} ]$, where $\epsilon$ and $\mu$ are the relative permittivity and permeability of the medium. The nontrivial bulk topology of $\mathbb{Z}_{2}$ TIs is well described by adding a term of the form $\mathcal{L} _{\theta} = (\alpha /  \pi Z _{0}) \, \theta  \boldsymbol{E} \cdot \boldsymbol{B}$, where $\alpha$ is the fine structure constant, $Z _{0} = \sqrt{\mu _{0}/\epsilon _{0}}$ is the impedance of free space, and $\theta$ is the topological magnetoelectric polarizability \cite{Qi-TFT}. For TIs, the only nonzero value compatible with TR symmetry is $\theta = \pi$ (modulo $2 \pi$), and thus has no effect on Maxwell equations in the bulk. Its only physical manifestation, a half-quantised Hall effect on the sample’s surfaces, becomes manifest only in the presence of surface magnetization, in which case we have $\theta = \pm (2n+1) \pi$, where $n \in \mathbb{Z} ^{+}$. Due to the existence of the $\theta$-term, the electromagnetic response of a fully gapped TI is described by Maxwell equations in matter, with however the modified constituent equations
\begin{align}
    \boldsymbol{D} = \epsilon \epsilon _{0} \boldsymbol{E} + \frac{\alpha \theta}{\pi Z _{0}} \boldsymbol{B} , \qquad  \boldsymbol{H} = \frac{1}{\mu \mu _{0}} \boldsymbol{B} - \frac{\alpha \theta}{\pi Z _{0}} \boldsymbol{E} .
\end{align}
The most important consequence of this theory is the TME effect to characterize the nontrivial $\mathbb{Z}_{2}$ topology of a TI in its electromagnetic properties. Specific TMEs have been predicted on the basis of this theory \cite{Qi-monopole, Karch, MCU1}, but it has been experimentally verified only through the measurement of Kerr and Faraday angles at the surface of a strained HgTe 3D TI \cite{Dziom}.

{\it Spherical TI in a monochromatic magnetic field.} Now let us consider a spherical nonmagnetic TI embedded in a dielectric fluid, as shown in figure 1. The system is subject to an externally applied monochromatic magnetic field $\boldsymbol{B} (\boldsymbol{r} , t) = \mbox{Re} \big\{\ \!\!\! \boldsymbol{B} _{0} e ^{-i \omega t} \big\}\ \!\!$, where $\boldsymbol{B} _{0} = \sum _{i} B _{0i} \boldsymbol{e} _{i}$ is a constant vector and $\mbox{Re} \big\{\ \big\}\ \!\!$ indicates the real part. The TI, of radius $R$, is characterized by the dielectric permittivity $\epsilon _{1} (\omega)$, magnetic permeability $\mu _{1} \approx 1$, and topological magnetoelectric polarizability $\theta$; while the dielectric fluid has dielectric function $\epsilon _{2} (\omega)$ and permeability function $\mu _{2} \approx 1$. The TI size is assumed to be small compared to the wavelength of the applied field, so we can make the time-harmonic approximation. That is, the electromagnetic fields can be written as $\boldsymbol{E} (\boldsymbol{r} , t) = \mbox{Re} \big\{\ \!\! \boldsymbol{E} (\boldsymbol{r}) e ^{- i \omega t} \big\}\ \!\!$ and $\boldsymbol{B} (\boldsymbol{r} , t) = \mbox{Re} \big\{\ \!\! \boldsymbol{B} (\boldsymbol{r}) e ^{- i \omega t} \big\}\ \!\!$, where $\boldsymbol{B} (\boldsymbol{r})$ and $\boldsymbol{E} (\boldsymbol{r})$ are the fields associated with the solution of the static Maxwell equations. This means that the electromagnetic field distribution in the quasistatic approximation only needs the EM fields in a constant magnetic field. Due to the TME effect, the magnetic field polarises the TI sphere, which can be described by the electric field
\begin{align}
    \boldsymbol{\mathcal{E}} (\boldsymbol{r} , \omega) = - \sum_{i}  \frac{3\tilde{\alpha}c}{3 ( 2 \epsilon _{2} + \epsilon _{1} ) + 2 \tilde{\alpha}^{2}}\, B_{0i}\,\boldsymbol{\mathcal{G}} _{i} (\boldsymbol{r}) \, , \label{E-field}
\end{align}
where $\tilde{\alpha} = \alpha (\theta / \pi)$, $c = (\epsilon _{0} \mu _{0}) ^{-1/2}$ is the speed of light, and
\begin{align}
    \boldsymbol{\mathcal{G}} _{i} (\boldsymbol{r}) = \left\lbrace  \begin{array}{c} \boldsymbol{e} _{i} \\[5pt] - \frac{R ^{3}}{r ^{3}} \left[3\,(\boldsymbol{e} _{r} \cdot \boldsymbol{e} _{i})\, \boldsymbol{e} _{r} - \boldsymbol{e} _{i} \right]    \end{array} \right.  \begin{array}{c} r<R \\[5pt] r>R  \end{array} . \label{G-func}
\end{align}
Clearly the electric field inside the TI is uniform as that of a uniformly polarised sphere, whilst the electric field outside is dipolar. Note that this electric field is a direct manifestation of the nontrivial topological order of the TI. Also, one can show that the magnetic field can be written as the externally applied magnetic field $\boldsymbol{B} _{0}$ plus a dipolar field of the form
\begin{align}
   \boldsymbol{\mathcal{B}} (\boldsymbol{r} , \omega) = \sum_{i} \frac{ \tilde{\alpha} ^{2}}{3 (2 \epsilon _{2} + \epsilon _{1} ) + 2 \tilde{\alpha} ^{2}} B _{0i} \,\xi(r)\, \boldsymbol{\mathcal{G}} _{i} (\boldsymbol{r}) , \label{B-field}
\end{align}
where $\xi (r) = 1$ for $r>R$ and $\xi (r) = -2$ for $r<R$.
\begin{figure}
\begin{center}
\includegraphics[scale=0.35]{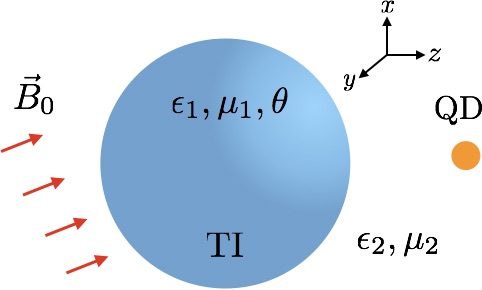}
\caption{\small Schematic of the topological insulator-quantum dot hybrid in the presence of an external magnetic field.}
\label{TI-QD-Fig}
\end{center}
\end{figure}
The time-dependent electromagnetic fields are obtained by Fourier transforming the previous results. To this end, a model for the dielectric function is necessary. Because of the low concentration of free carriers in insulators, the dielectric function $\epsilon _{1} (\omega)$ can be well modeled by
\begin{align}
    \epsilon _{1} (\omega) = 1 + 
\frac{\omega_{e} ^{2}}{\omega_{R} ^{2}-\omega(\omega + i \gamma _{0})} \, , \label{Dielectric-function}
\end{align}
with $\omega _{R}$ and $\omega _{e}$ the resonant and natural frequencies, respectively. The damping parameter $\gamma _{0}$ (for which $\gamma _{0} \ll \omega _{R}$) accounts for energy dissipation due to ohmic losses in the TI. Hence, taking a constant value for $\epsilon _{2}$ and using the dielectric function (\ref{Dielectric-function}) we find that when $\gamma _{0} \ll \omega$ the electromagnetic fields (\ref{E-field}) and (\ref{B-field}) can be approximated as
\begin{align}
    \boldsymbol{\mathcal{E}} (\boldsymbol{r} , \omega) &= - \eta \frac{\omega _{0} ^{2} / 2 \Omega}{\omega - \Omega + i \gamma _{0} / 2} \sum_{i} c \, B _{0i}\,\boldsymbol{\mathcal{G}} _{i} (\boldsymbol{r}) ,   \label{E-field2} \\ \boldsymbol{\mathcal{B}} (\boldsymbol{r} , \omega) &=  - (\tilde{\alpha} / 3c) \xi(r) \boldsymbol{\mathcal{E}} (\boldsymbol{r} , \omega) , \label{B-field2}
\end{align}
where $\Omega = \sqrt{\omega _{R} ^{2} + \omega _{0} ^{2}}$, $\omega _{0} = \omega _{e} \sqrt{\eta / \tilde{\alpha}}$ and
\begin{align}
    \eta = \frac{3 \tilde{\alpha}}{3 ( 2 \epsilon_{2} + 1 ) + 2 \tilde{\alpha}^{2}} . \label{Eta}
\end{align}
Evidently, in this limit the electromagnetic fields (\ref{E-field2}) and (\ref{B-field2}) follow Lorentzian spectra, whose approximation is appropriate when the TI is interacting with a dipole whose resonant frequency is close to plasmon resonance.

{\it Localised solutions.} The EM fields derived above are not localised solutions because they are driven by a monochromatic plane wave that extends infinitely in space. However, localised solutions are needed in order to quantise the field modes at the TI surface. To this end, we have to excite the TI with an impulse function rather than by a monochromatic field, since after the impulse has ended, only the localised modes will remain. For an input field of the form $B _{0} (t) = B _{0} \, \delta (t)$, the electric field in the time domain is obtained by Fourier transforming Eq. (\ref{E-field2}). The result is
\begin{align}
    \boldsymbol{\mathcal{E}} (\boldsymbol{r} , t) &= \sum _{i} \Lambda _{i} \, \sin ( \Omega t) \,  e^{-\gamma_{0} t / 2} \, \boldsymbol{\mathcal{G}} _{i} (\boldsymbol{r}) \, , \label{E-Field-Localised}
\end{align}
where $\Lambda_{i} = - \eta \, (\omega _{0} ^{2} / 2 \Omega )\, B _{0i} c$. Clearly, the identity of  Eq. (\ref{B-field2}) for the magnetic field remains in the time domain. By direct calculation one can confirm that these fields represent localised solutions to the field equations: $\boldsymbol{\mathcal{G}} _{i}$ satisfies both the field equations and the boundary conditions at $\omega = \Omega$ in the undamped limit.

\section{III. Quantum optical model} \

{\it Quantisation of the TI response.} At this stage we ignore for the moment the term $\gamma _{0}$, such that the fields are steady-state sinusoidal functions. Such term will be included later with the incorporation of a continuum of reservoir modes.  

From Eq. (\ref{E-Field-Localised}) it is clear that the different cartesian components of the electromagnetic fields are orthogonal. Hence we may quantise each individually. A further simplification is in order. Due to the smallness of the TME effect of TIs, in the following we will retain only the lowest order contribution in the fine structure constant. Let us consider the energy of the $i$th mode of the EM fields in a dispersive media: $U _{i} = \frac{\epsilon _{0}}{2} \int [ \boldsymbol{\mathcal{E}} ^{2} _{i} \frac{d (\omega \epsilon)}{d \omega}  + (1 / \mu ) \boldsymbol{\mathcal{B}} ^{2} _{i}] \, d ^{3} \boldsymbol{r}$. Since the magnetic contribution is suppressed by a factor of $\alpha ^{2} \! \sim \! 10 ^{-5}$ with respect to that of the electric field, the energy of the $i$th mode can be approximated as
\begin{align}
    U _{i} &= \frac{\epsilon _{0}}{2} \Lambda _{i} ^{2} \sin ^{2} ( \Omega t) \int \vert \boldsymbol{\mathcal{G}} _{i} (\boldsymbol{r}) \vert ^{2} \,  \frac{d [ \mbox{Re}( \omega \epsilon ) ]}{d \omega} \bigg| _{\omega = \Omega} \, d ^{3} \boldsymbol{r} .  \label{Energy-ithMode}
\end{align}
Field quantisation is followed by definition of the normalised amplitude $A _{i} = \Lambda _{i} / \mathcal{N}$, where
\begin{align}
   \frac{1}{\mathcal{N} ^{2}} = \frac{\epsilon _{0}}{2 \hbar \Omega} \int \vert \boldsymbol{\mathcal{G}} _{i} (\boldsymbol{r}) \vert ^{2} \, \frac{d [ \mbox{Re}( \omega \epsilon ) ]}{d \omega} \Bigg| _{\omega = \Omega} \, d ^{3} \boldsymbol{r} , \label{Normalization}
\end{align}
such that the energy stored in the electromagnetic fields (\ref{Energy-ithMode}) can be rewritten in the simple form $U _{i} = \hbar \Omega A _{i} ^{2} \sin ^{2} ( \Omega t)$. Clearly, energy conservation requires another form of energy due to current flowing in the TI surface. To this end, we add a second term in the Hamiltonian accounting for the periodic conversion between stored potential energy (represented by the energy of the field $U _{i}$) and kinetic energy due to current flowing in the TI. This energy must be of the form $K _{i} = \hbar \Omega A _{i} ^{2} \cos ^{2} ( \Omega t)$ such that the total energy $H _{i} = U _{i} + K _{i}$ is constant at all times.

The above analysis suggests that introducing the time-dependent amplitude $\mathcal{A} _{i} (t) = A _{i} \sin (\Omega t)$, the total Hamiltonian of the field modes can be written as $H _{i} = (\hbar / \Omega ) \big( \dot{\mathcal{A}} _{i} ^{2} + \Omega ^{2} \mathcal{A} _{i} ^{2} \big)$, which is similar to that of a one-dimensional harmonic oscillator. The two variables $\mathcal{A}_{i}$ and $2\hbar\dot{\mathcal{A}}_{i}/\Omega$ form a pair of canonical conjugate variables that can be quantised. To this end, we promote them to quantum operators as $\mathcal{A} _{i} \to \hat{x} _{i}$ and $2 \hbar \dot{\mathcal{A}} _{i} / \Omega \to \hat{p} _{i}$, which satisfy the commutation relation $\left[\hat{x} _{i},\,\hat{p}_{j}\right]=i\delta_{ij}\hbar$. Now, introducing the bosonic creation an annhilation operators, $\hat{a} _{i} = \hat{x} _{i} + ( i/ 2 \hbar ) \hat{p} _{i}$ and $\hat{a} _{i} ^{\dagger} = \hat{x} _{i} - ( i/ 2 \hbar ) \hat{p} _{i}$, the quantum Hamiltonian describing the surface TI modes becomes
\begin{align}
    \hat{H}_{\mathrm{TI}} = \hbar \Omega \sum _{i} \Big( \hat{a} ^{\dagger} _{i} \hat{a} _{i} + 1/2 \Big). \label{HamiltonianTI}
\end{align}
This program also implies the quantisation of the electromagnetic fields over the TI surface. Indeed, one can further show that $\Lambda _{i} \to \frac{\mathcal{N}}{2} \big( \hat{a} _{i} + \hat{a} _{i} ^{\dagger} \big)$. Therefore, in the steady-state condition, the electric field can be quantised as 
\begin{align}
    \boldsymbol{\mathcal{E}} (\boldsymbol{r} , t) &= \sqrt{\frac{\hbar \Omega}{2 \epsilon _{0} V _{m}}} \sum _{i} \big( \hat{a} _{i} + \hat{a} _{i} ^{\dagger} \big) \, \boldsymbol{\mathcal{Y}} _{i} (\boldsymbol{r}) \, , \label{E-Field-Quantised}
\end{align}
where the mode volume $V_{m}$ is defined as the ratio between the total energy to the energy density inside the TI, i.e.
\begin{align}
    V _{m} = \frac{\int \vert \boldsymbol{\mathcal{G}} _{i} (\boldsymbol{r}) \vert ^{2} \, \frac{d\mbox{\small Re}( \omega \epsilon )}{d \omega} \Big| _{\omega = \Omega} \, d ^{3} \boldsymbol{r}}{\mathcal{U}_{0}} ,
\end{align}
with
\begin{equation}
\mathcal{U}_{0} = 
\vert \boldsymbol{\mathcal{G}} _{i} (\boldsymbol{0}) \vert ^{2} \, \frac{d \mbox{Re}( \omega \epsilon _{1} )}{d \omega} \Big| _{\omega = \Omega}\,. 
\end{equation}
Finally, $\boldsymbol{\mathcal{Y}} _{i} (\boldsymbol{r}) = \boldsymbol{\mathcal{G}} _{i} (\boldsymbol{r}) / \sqrt{\mathcal{U}_{0}} $, which is a rescaled version of $\boldsymbol{\mathcal{G}} _{i} (\boldsymbol{r})$. Note that the quantised electric field in Eq. (\ref{E-Field-Quantised}) depends on the electric field per photon $\mathcal{E} _{0} = \sqrt{\hbar \Omega / 2 \epsilon _{0} V _{m} }$, but not on the strength of the impulse magnetic field $B _{0}$.\\ 

{\it Hamiltonian of the system.} When the distance between the TI and the QD is large as compared with the TI radius, the dipolar approximation provides a reasonable estimate for the interaction, i.e. $\hat{H}_{\mathrm{int}} = - \hat{{\bold p}} \cdot \boldsymbol{\mathcal{E}}$, where $\hat{{\bold p}}$ is the dipole operator and $\boldsymbol{\mathcal{E}}$ is the quantised electric field derived above. When the QD is near the TI surface, higher-order multipole moments will be required to properly characterise the interaction.

For simplicity we shall consider a spherically symmetric quantum dot, such that the matrix elements of the dipole operator, ${\bold p} _{nm} = \langle m \vert \hat{{\bold p}} \vert n \rangle $, points along a specific direction. In this manner, $\hat{H}_{\mathrm{int}}$ couples field operators (dipole and electric field) pointing along the same direction. Applying the two-level approximation \cite{Landau, Frasca}, the dipole operator takes the form $\hat{{\bold p}} = d (\hat{\sigma} ^{+} + \hat{\sigma} ^{-}) \boldsymbol{e} _{i}$ (with $i=x,y,z$), where $d$ is the dipole moment of the transition and $\hat{\sigma} ^{+}$ and $\hat{\sigma} ^{-}$ are the Pauli raising and lowering operators respectively. Therefore, the interaction Hamiltonian for a single mode interacting with the QD can be expressed as
\begin{align}
    \hat{H}_{\mathrm{int}} = \hbar g (r) \, (\hat{\sigma} ^{+} \hat{a} + \hat{\sigma} ^{-} \hat{a} ^{\dagger} ) , 
\end{align}
where we have taken only the energy preserving contributions. The TI-QD coupling strength depends on whether, the electric field points along the dipole direction (longitudinal coupling, LC) or in the transverse direction (transverse coupling, TC). Explicitly, the coupling strength reads
\begin{align}
   \!\! g(r) = \left\lbrace \begin{array}{l} \!\!\!
    +2 \, \frac{d}{\hbar} \sqrt{\frac{\hbar \Omega}{2 \epsilon _{0} V _{m} \mathcal{U} _{0}}} \frac{R ^{3}}{r ^{3}}   \\[8pt]     \!\!\! -1 \, \frac{d}{\hbar} \sqrt{\frac{\hbar \Omega}{2 \epsilon _{0} V _{m} \mathcal{U} _{0}}} \frac{R ^{3}}{r ^{3}}     \end{array} \right. \begin{array}{l} \!\!\!      \mbox{LC}   \\[8pt]  \!\!\!      \mbox{TC}  
    \end{array} . \label{CouplingStrength}
\end{align}
This result implies that the longitudinal coupling is preferable for experimental detection since it is twice stronger than the transverse coupling. All in all the Hamiltonian describing the closed QD-TI system is simply $\hat{H} = \hat{H}_{\mathrm{TI}} + \hat{H}_{\mathrm{dip}} + \hat{H}_{\mathrm{int}}$, where $\hat{H}_{\mathrm{dip}} = \hbar \omega _{a} \hat{\sigma} ^{+} \hat{\sigma} ^{-}$ is the dipole Hamiltonian (being $\omega _{a}$ the resonant frequency of the dipole). 

Damping effects due to the interaction of the system with the environment has to be included in any realistic model.  In this Letter we consider that the system is coupled with a continuum reservoir of radiative output modes and a reservoir of phonon modes. The Hamiltonian of the radiative and phonon reservoirs is given by \cite{Scully, Carmichael}
\begin{align}
\hat{H} _{\mbox{\scriptsize B}} =\int \hbar \omega ( \hat{b} ^{\dag} {}_{\!\! \omega} \hat{b}_{\omega} + \hat{c} ^{\dag} {}_{\!\! \omega} \hat{c}_{\omega } ) d\omega , \label{ec99a}
\end{align}
while the Hamiltonian describing the interaction between the system and the reservoirs is
\begin{align}
\hat{H}_{\mbox{\scriptsize SB}} & = i\hbar\int ( T_{1} \hat{b} ^{\dag} {}_{\!\! \omega} \hat{a} + T _{2} \hat{c} ^{\dag} {}_{\!\! \omega} \hat{a} + T _{3} \hat{b} ^{\dag} {}_{\!\! \omega} \hat{\sigma} ^{-} ) d \omega + \mbox{h.c.} \label{ec99b}
\end{align}
Here, $\hat{b} ^{\dag} {}_{\!\! \omega}$ and $\hat{b} _{\omega}$ ($\hat{c} ^{\dag} {}_{\!\! \omega}$ and $\hat{c} _{\omega}$) are the creation and annihilation operators corresponding to the radiative (phonon) modes. The terms $T_{1}=\sqrt{\gamma_{r}/2\pi}$ and $T_{2}=\sqrt{\gamma_{0}/2\pi}$ represent the coupling strength between the TI and the reservoir modes, while $T_{3}=\sqrt{\gamma_{s}/2\pi}$ represents the coupling strength between the dipole and the radiative modes. Here, $\gamma_{r}$, $\gamma_{0}$ and $\gamma_{s}$ are the scattering rate into free-space modes, the energy dissipation due to ohmic losses, and the spontaneous emission rate of the dipole, respectively. Putting it all together, the full Hamiltonian of the open system is
\begin{equation}
    \hat{H} = \hat{H} _{\mbox{\scriptsize TI}} + \hat{H} _{\mbox{\scriptsize dip}} + \hat{H} _{\mbox{\scriptsize int}} + \hat{H} _{\mbox{\scriptsize B}} + \hat{H}_{\mbox{\scriptsize SB}} . \label{FullHamiltonian}
\end{equation}

\section{IV. Absorption spectrum}

It is well-known that the optical absorption spectra is given by the Fermi's golden rule according to
\begin{align}
    \sigma(\omega)\propto \sum _{f}
\vert \langle f;n-1 \vert \hat{V} \vert i;n \rangle \vert^{2} \, \delta( \omega _{fi} - \omega ) , \label{AbsSpectrum}
\end{align}
where $\left| i \right>$ represents the initial ground state of the system, $\left| f \right>$ is the corresponding final state, $\omega _{fi}$ is the frequency associated to the energy difference between the final and initial states, and $n$ is the number of external photons with frequency $\omega$. For the TI-QD hybrid under consideration, the perturbation Hamiltonian has the generic form $\hat{V} \propto \hat{A} \hat{a} ^{\dagger} + \hat{A} ^{\dagger} \hat{a}$, where $\hat{a}$ and $\hat{A}$ ($\hat{a}^{\dagger}$ and $\hat{A}^{\dagger}$) are the annihilation (creation) operators for the external photons and excitations of the system, respectively. In this manner, $\hat{A}$ connects the initial and final states of the system, thus governing the optical absorption properties.

As shown in Ref. \cite{manjavacas}, the optical absorption spectrum (\ref{AbsSpectrum}) can be expressed in terms of the retarded Zubarev's Green function by
\begin{align}
     \sigma(\omega) \propto - \mbox{Im} \; \langle \langle \hat{A} ; \hat{A} ^{\dagger} \rangle \rangle _{\omega + i 0 ^{+}}  . \label{AbsSpectrum2}
\end{align}
We recall the definition of the retarded Zubarev's Green function of two operators $\hat{F}$ and $\hat{G}$ in the frequency domain \cite{Zubarev}:
\begin{align}
    \langle \langle \hat{F} ; \hat{G} \rangle \rangle _{\omega+i0^{+}} = \frac{1}{i \hbar} \int _{0} ^{\infty} dt e ^{i(\omega + i 0 ^{+}) t} \theta (t) \langle [\hat{F} (t), \hat{G} (0)] _{\eta} \rangle , \label{GZ-function}
\end{align}
where $\hat{F} (t)$ means the Heisenberg representation, $\theta (x)$ is the usual step function and the brackets $[\hat{F}, \hat{G}] _{\eta} = \hat{F} \hat{G} - \eta \hat{G} \hat{F}$ stands for the commutator (anticommutator) of bosonic (fermionic) operators for $\eta = 1$ ($\eta = -1$). In this way, by computing the retarded Zubarev's Green function $\langle \langle \hat{A} ; \hat{A} ^{\dagger} \rangle \rangle$ we will immediately obtain the optical absorption spectrum. To this end, we have to employ its equation of motion \cite{Zubarev}
\begin{align}
    \hbar \omega \langle \langle \hat{A} ; \hat{A} ^{\dagger} \rangle \rangle = \langle[\hat{A} , \hat{A} ^{\dagger}] _{\eta} \rangle + \langle \langle [\hat{A} , \hat{H} ] ; \hat{A} ^{\dagger} \rangle \rangle  \label{EqMotion} ,
\end{align}
where $\hat{H}$ is the Hamiltonian of the system. As we can see, this expression depends on another Zubarev's Green function $\langle \langle [\hat{A} , \hat{H} ] ; \hat{A} ^{\dagger} \rangle \rangle$, which can also be calculated by writing down its equation of motion. Iterating this process, one obtains a hierarchy of equations that may need to be truncated at some point by applying a physical approximation. This program will produce a linear system of equations from which we will obtain $\langle \langle \hat{A} ; \hat{A} ^{\dagger} \rangle \rangle$ and hence the optical absorption spectrum. The calculation is simple, but not straightforward. In fact it is too long to be shown here. Therefore we will present only the final result. We truncate the iteration process by approximating the operator $\hat{\sigma} ^{+}\hat{\sigma} ^{-}$ by its expectation value $\langle \hat{\sigma} ^{+}\hat{\sigma} ^{-} \rangle = n$. The result is
\begin{align}
    \sigma (\omega) & \propto \mbox{Im} \bigg\{\ \omega - \Omega - \lambda _{11} - \lambda _{22} \notag \\ & \hspace{1cm} - \frac{(1-2n) [g(r) + \lambda _{13}] [g(r) + \lambda _{31}] }{\omega - \omega _{a} - (1-2n) \lambda _{33}} {\bigg\}\ }^{\! -1} , \label{AbsEspectrumFin}
\end{align}
where we have defined 
\begin{align}
    \lambda _{ij} (\omega) &= \int \frac{T _{i} ^{\ast} (\omega ^{\prime}) T _{j} (\omega ^{\prime})}{\omega - \omega ^{\prime} - i 0 ^{+}} d \omega ^{\prime} \notag \\ &= \mbox{p.v.} \int \frac{T _{i} ^{\ast} (\omega ^{\prime}) T _{j} (\omega ^{\prime})}{\omega - \omega ^{\prime}} d \omega ^{\prime} + i \pi T _{i} ^{\ast} (\omega) T _{j} (\omega) . \label{NewFrequencies}
\end{align}
Here, p.v. stands for the Cauchy principal value. As expected, the resonance frequency of the quantum modes on the TI is modified by the interaction with the QD.

In general, plasmonic nanoparticles can interact with light in several different ways. For example, the scattering, absorption and extinction spectra depends on the nanoparticle size, shape, and composition. In many applications involving the use of electromagnetic fields, as the one tackled in this Letter, specific frequencies are used to excite the plasmonic nanoparticles in order to obtain the strongest output field. As shown in Ref. \cite{Near}, for systems with a size of a few nanometers, the field strength trends with absorption and not extinction or scattering. Therefore, the optical response of our system is well described by absorption.

\section{V. Numerical results} Here we apply our results to a realistic TI-QD hybrid. To this end we use appropriate values for the parameters in Eq. (\ref{AbsEspectrumFin}). Let us  consider a CdSe QD interacting with a TI spherical nanoparticle made of the TI TlBiSe$_{2}$. The parameters for this TI has found to be $\mu _{1} = 1$ and $\epsilon_{1}(0)=1+(\omega_{e}/\omega_{R})\sim 4$ \cite{siapkas}. We take an energy for the TI of  $\hbar\Omega = 2.0$ eV and a scattering rate into free-space modes $\gamma_{r}\sim 1.5\times 10^{-4}$ eV. Further, the damping parameter $\gamma_{0}$ satisfies the condition $\gamma_{0}\ll\omega_{R}$, hence it plays a secondary role and we can safely neglect it in the numerical simulations. The TI spherical nanoparticle is assumed to have a radius of $R\sim 4$ nm and be embedded in a polymer layer of poly(methyl methacrylate) with permittivity $\epsilon_{2}=1.5$ \cite{waks}. On the other hand, we suppose a CdSe QD with a size of 4 nm and with a spontaneous emission decay rate $\gamma_{s} = 6.5\times 10^{-8}$ eV \cite{waks}. The QD resonance energy $\hbar\omega_{a}$ is within a range of $2.7-1.3$ eV, an appropriate energy range for CdSe QDs. Also, we set $d=6.4\times 10^{-28}$ Cm for the transition dipole moment \cite{indian}. Finally, we consider an output magnetic field with a wavelength of $\lambda=620$ nm and occupation number $n=0.2$ \cite{nanoscale2018}.

\begin{figure*}
\vspace{-0.3cm}
\begin{center}
\includegraphics[scale=0.44]{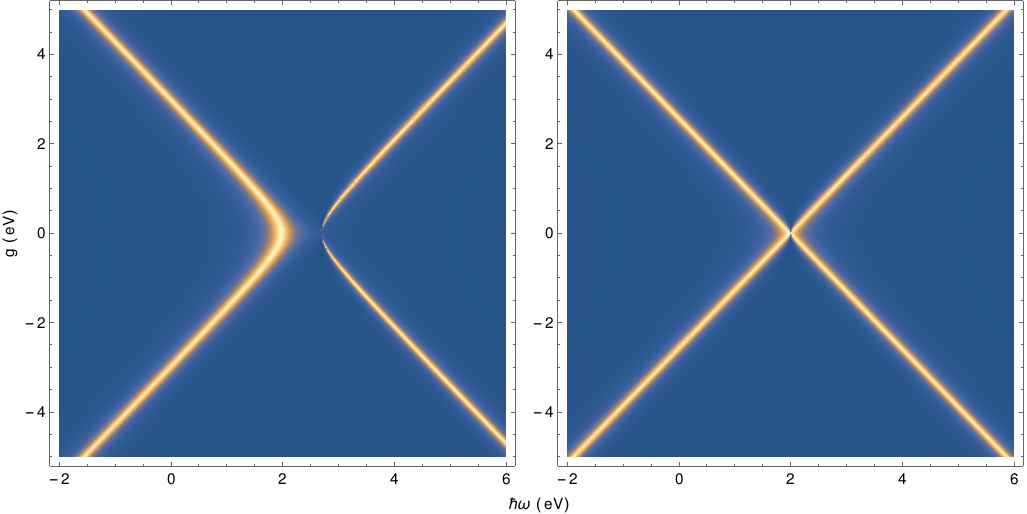}
\caption{Density plot of the optical absorption spectrum of the TI-QD hybrid as a function of the coupling strength ($g$) and the energy $\hbar \omega$ of the photonic field,  for longitudinal (left) and transverse (right) coupling. The TI-QD Fano resonance achieve its maximum approach for the QD resonance energy $\hbar\omega_{a}=2.0$ eV.}
\label{esp1}
\end{center}
\end{figure*}

\begin{figure*}
\vspace{-0.5cm}
\begin{center}
\includegraphics[scale=0.45]{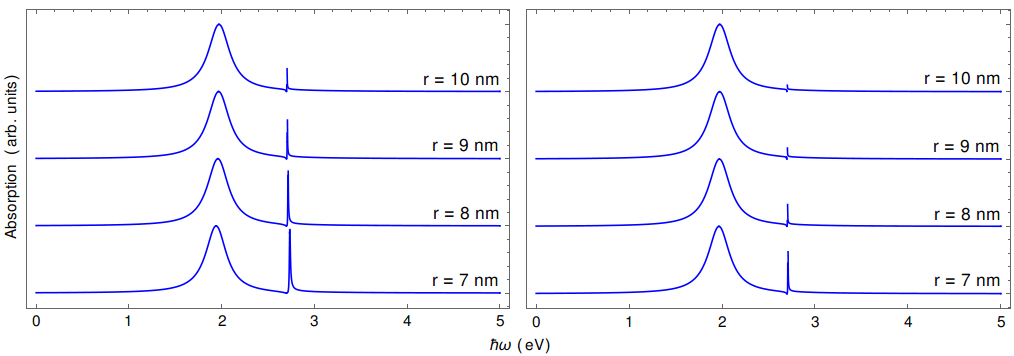}
\caption{Optical absorption spectrum of the TI-QD hybrid for longitudinal (left) and transverse (right) coupling for the distances $r = 7,\,8,\,9,\,10$ nm. We fix the QD resonance energy $\hbar\omega_{a}=2.7$ eV.}
\label{esp2}
\end{center}
\end{figure*}

\begin{figure*}
\vspace{-0.5cm}
\begin{center}
\includegraphics[scale=0.44]{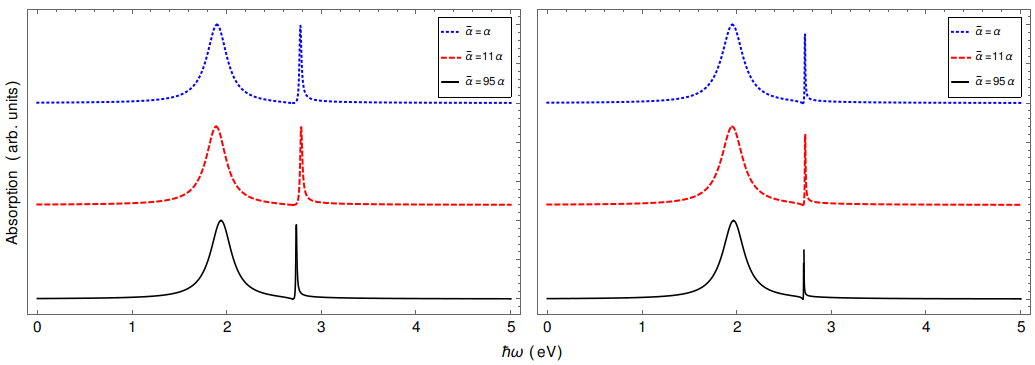}
\caption{Optical absorption spectrum of the TI-QD hybrid for longitudinal (left) and transverse (right) coupling for the rescaled topological magnetoelectric  polarizabilities $\tilde{\alpha}=\alpha$ (dotted blue line), $\tilde{\alpha}=11\alpha$ (dashed red line), and $\tilde{\alpha}=95\alpha$ (continuous black line). Here we fix $\hbar\omega_{a}=2.7$ eV and $r=7$ nm.}
\label{esp3}
\end{center}
\end{figure*}

Figure \ref{esp1} shows the optical absorption spectrum of the system for LC (at left) and TC (at right), for a fixed value of $\hbar\omega_{a}=2.0$ eV. Clearly, we observe asymmetric line shapes which we identify as Fano resonances. Overall, Fano resonances results from the interaction between a continuum of modes and a discrete mode \cite{Limonov}. In our work, we identify the continuum modes as the output radiative and phonon modes reservoirs, whilst the discrete mode correspond to the resonant coupling of the TI-QD hybrid. Of course, the origin of our Fano resonances lies only in the topological magnetoelectric effect of the TI nanoparticle (as the induced electromagnetic fields suggest), and thus they are a direct manifestation of the $\mathbb{Z}_{2}$ invariant of TIs. In other words, Fano resonances disappear for a vanishing topological magnetoelectric polarizability $\tilde{\alpha}$. Note that the distance between the peaks is related with the factor appearing in the coupling strength (\ref{CouplingStrength}), i.e. for TC the peaks are closer than for LC. Also, by taking $\hbar\omega_{a}=2.7$ eV, in Fig. \ref{esp2} we present the optical absorption spectrum for different values of the TI-QD distance $r$. As expected, the peak associated with the QD (the narrow one) becomes more pronounced as the TI-QD distance is reduced, and decreases faster for TC. Finally, in Fig. \ref{esp3} we illustrate the optical absorption spectrum as a function of the (rescaled) topological magnetoelectric polarizability $\tilde{\alpha}$, for fixed values of $\hbar\omega_{a}=2.7$ eV and $r=7$ nm. In particular, we take $\tilde{\alpha}=\alpha$ (which is the lowest nontrivial possible value), $\tilde{\alpha}=11\,\alpha$ and $\tilde{\alpha}=95\,\alpha$. Here, we observe that for TC (at right) the Fano resonances approach subtly faster than for LC (at left). Outstandingly, the Fano resonances increase faster for TC than for LC.

\section{VI. Conclusion and discussion} In this Letter we have shown that a {\it topological hybrid plexciton} formed by a quantum dot and a TI nanoparticle subject to an impulse magnetic field exhibits Fano resonances which are direct manifestation of the $\mathbb{Z}_{2}$ invariant of TIs. We applied our results to a realistic TI-QD hybrid composed by a nanosphere made of TlBiSe$_2$ interacting with a CdSe QD in a polymer layer such as poly(methyl methacrylate). Our results can also be tested with linear magnetoelectrics materials such as Cr$_2$O$_3$ and in some multiferroics. However, as we discussed, for TIs with TR invariance in the bulk, the TME effect with a quantised value of $\tilde{\alpha}$ is a unique signature of the topological nontriviality of the band structure.

The term {\it topological Fano resonance} has been introduced recently to name ultrasharp asymmetric line shapes which are protected against geometrical disorder of the sample \cite{Zangeneh}. Of course, the topological Fano resonances we report here do not belong to this classification, since the topological protection here  is related with the band structure of the TI, and not directly with its geometrical form.

Furthermore, the theoretical analysis carried out in this work can be extend to other topological materials, such as topological Weyl semimetals and Dirac semimetals. In that case, the metallic character of the sample would enhance the interaction with a quantum dot, but at the same time, the topological contribution could be overwhelmed by the nontopological parts. It is worth to mention that the Fano resonances reported in this paper are direct manifestations of the magnetoelectric polarizability $\tilde{\alpha}$ characteristic of topological insulators.

\acknowledgments
A.M.-R. and L.C.-E acknowledges support from DGAPA-UNAM project IA101320.

\end{document}